\documentclass[runningheads]{llncs}
\usepackage{graphicx}
\usepackage{amssymb}
\usepackage{amsmath}
\usepackage{booktabs}
\usepackage{tabularx}
\usepackage{multirow}
\usepackage{caption}
\usepackage{subcaption}
\usepackage{amsfonts}
\newcommand*\samethanks[1][\value{footnote}]{\footnotemark[#1]}

\DeclareMathOperator{\softmax}{softmax}

\begin{document}
\title{Dementia Severity Classification under Small Sample Size and Weak Supervision in Thick Slice MRI}
%
%
\author{Reza Shirkavand\thanks{Equal contribution}\inst{1}, Sana Ayromlou\samethanks{}\inst{1}, Soroush Farghadani\samethanks{}\inst{1}, Maedeh-sadat Tahaei\inst{1}, Fattane Pourakpour\inst{2}, Bahareh Siahlou\inst{2}, Zeynab Khodakarami\inst{2}, Mohammad H. Rohban\inst{1}, Mansoor Fatehi\inst{2,3}, \and Hamid R. Rabiee\inst{1}}
%
\authorrunning{Shirkavand et al.}
%
\institute{
Sharif University of Technology, Tehran, Iran.\\
\and
Iranian Brain Mapping Biobank, National Brain Mapping Laboratory, Tehran, Iran.\\
\and Virtual University of Medical Sciences, Tehran, Iran.\\
\email{rohban@sharif.edu}}
\maketitle              

\begin{abstract}
Early detection of dementia through specific biomarkers in MR images plays a critical role in developing support strategies proactively. Fazekas scale facilitates an accurate quantitative assessment of the severity of white matter lesions and hence the disease. Imaging Biomarkers of dementia are multiple and comprehensive documentation of them is time-consuming. Therefore, any effort to automatically extract these biomarkers will be of clinical value while reducing inter-rater discrepancies. To tackle this problem, we propose to classify the disease severity based on the Fazekas scale through the visual biomarkers, namely the Periventricular White  Matter (PVWM) and the Deep White Matter (DWM) changes, in the real-world setting of thick-slice MRI. Small training sample size and weak supervision in form of assigning severity labels to the whole MRI stack are among the main challenges. To combat the mentioned issues, we have developed a deep learning pipeline that employs self-supervised representation learning,  multiple instance learning, and appropriate pre-processing steps. We use pretext tasks such as non-linear transformation, local shuffling, in- and out-painting for self-supervised learning of useful features in this domain. Furthermore, an attention model is used to determine the relevance of each MRI slice for predicting the Fazekas scale in an unsupervised manner. We show the significant superiority of our method in distinguishing different classes of dementia compared to state-of-the-art methods in our mentioned setting, which improves the macro averaged F1-score of state-of-the-art from 61\% to 76\% in PVWM, and from 58\% to 69.2\% in DWM.

\keywords{Dementia Early Detection  \and Fazekas scale \and Weak Supervision \and Self-Supervised Learning.}
\end{abstract}

\section{Introduction}

Dementia is a collective term that is used to describe a variety of syndromes in which there is a decline in cognitive functions. Nearly 50 million people are currently diagnosed with dementia worldwide, and the number is estimated to rise to 115 million by 2050 \cite{livingston2017dementia}. In addition to cognitive laboratory tests and medical records, clinicians use visual biomarkers to diagnose dementia. Several biomarkers of dementia, including Periventricular White Matter (PVWM) lesions and Deep White Matter (DWM) lesions, can be detected in the Magnetic Resonance Images (MRI) of the brain \cite{ikram2010brain}. These biomarkers can help to detect brain changes in subjects, irrespective of evident deterioration in their cognitive ability. Therefore, they are  crucial for early diagnosis, and effective planning of treatment and support strategies. Moreover, the Fazekas scale that is widely used for describing the severity of White Matter Lesions (WML) \cite{fazekas1987mr}, facilitates the quantitative assessment of dementia and provides insight into the progression of the disease and patient outcome.

Although manual classification of the Fazekas scale of PVWM and DWM lesions is the most accurate method, it is tedious and expensive and usually suffers from intra-observer and inter-observer variability. This warrants the need for developement of an automated method in classifying the disease state. 
Over the past two decades, several methods \cite{cuingnet2011automatic,farooq2017deep,hon2017towards,wang2018classification,gunawardena2017applying,taqi2018impact,islam2018brain,mendoza2020single,wu2018discrimination,liu2018landmark,qiu2018fusion,hosseini2016alzheimer,li2017alzheimer} have been proposed to help in diagnosing various forms of Dementia, including Mild Cognitive Impairment (MCI) and Alzheimer’s Disease (AD), based on the neuroimaging data . Here, our focus is on the deep neural networks as the earlier methods based on manually engineered features are concerned with sub-optimal performances.

There are two main approaches for classifying the disease based on MRIs through 2D convolutional neural networks (CNNs). These two directions are based on the input format of the 2D CNN, namely, slice-level and patch-level inputs. Along the former direction, some previous work used transfer learning, to tackle the issue of small training set, on models that are pre-trained on a set of natural images like ImageNet \cite{farooq2017deep}. 
However, some others designed their own simple CNNs \cite{wang2018classification,gunawardena2017applying,taqi2018impact}. One of the main drawbacks of this approach is the shortage of data to learn deep networks from scratch.

In the second set of approaches, extracting multiple patches from 2D slices provides a greater data size and lower dimensionality to address the mentioned issue \cite{islam2018brain,mendoza2020single}. In such methods, the ultimate decision for the subjects' label was based on the majority of labels that are associated with the patches. It is worth noting that the aforementioned problem about specified labels is even more severe in the patch-based methods. 
To utilize the spatial information in the third dimension, some studies focused on 3D approaches that take the whole MRI stack as input \cite{aderghal2018classification,liu2018landmark,hosseini2016alzheimer}. 
It should be noted that training of such methods requires large training sets, and the risk of overfitting is high \cite{li2017alzheimer}.  

In this work, we focus on two key challenges of small training set, and weak supervision of assigning disease progression labels to the entire MR stack of images. To tackle these issues, we propose a self-supervised learning scheme, and multiple instance learning (MIL), and show that such a combination would effectively enhance state-of-the-art by a large margin on a real-world dataset, which is generated by our team. In brief, our main contributions are the following:
\begin{enumerate}
    \item 
    To the best of our knowledge, this is the first study that targets classifying the severity of dementia based on the Fazekas scale of its visual biomarkers.
    
    \item
    We use thick-slice brain MRI data, an efficient solution that is mostly used in clinical practice, which is faster to acquire, reduces patient's exposure, and makes it easier for academic studies to be used in practice.
    
    
    \item
    We demonstrate our method's capability to discriminate the severity of dementia. The proposed framework outperforms the state-of-the-art in multi-class classification of the AD and the MCI, a similar setting to our problem. 
\end{enumerate}

\section{Method}
\begin{figure}[t]
	\centering
	\includegraphics[width=1\textwidth]{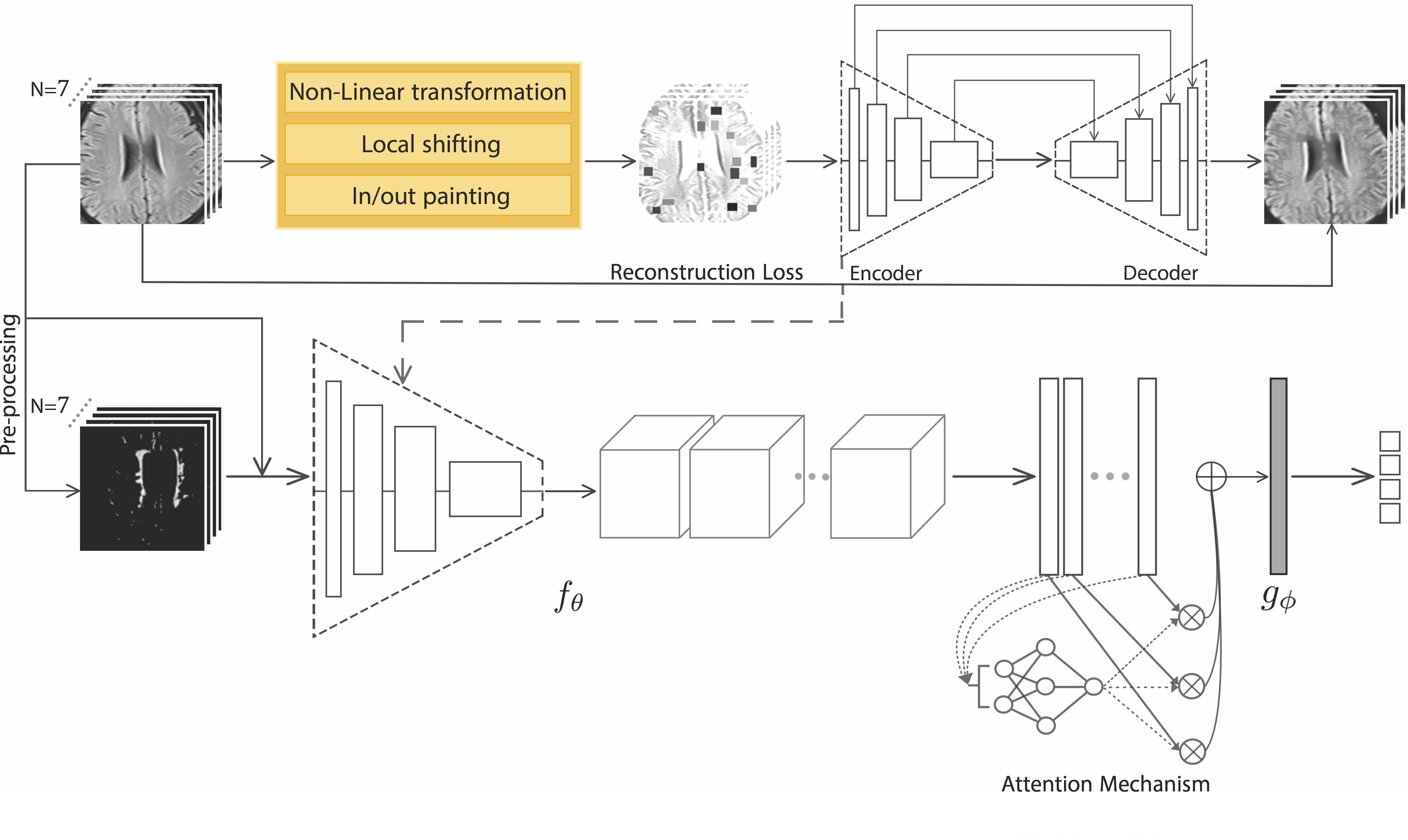}
	\caption{The framework of dementia biomarker classification in thick-slice 2D MR images. \textbf{Top:} a U-Net is first trained to reconstruct the original images from the transformed ones. \textbf{Bottom:} The encoder of the U-Net together with convolutional layers are used to extract features from each slice of both pre-processed and original MR images in the MIL component. The attention mechanism network produces a weighted average of the obtained features.}
	\label{network-diagram}
\end{figure}

Our proposed classification method, Fig. \ref{network-diagram}, consists of three main components: a data pre-processing step, self-supervised learning, and multiple instance learning, which we describe next. 

\subsection{Data Preprocessing}

As it is mentioned earlier, the classification task is mainly based on a specific type of white matter lesions. Hence, the CNN model, which can successfully classify patients based on the aforementioned biomarker, should mostly attend to the areas of the input image containing white matter lesions. Consequently, we designed a component for white matter lesion segmentation to be performed on the given MR images. This component is comprised of three steps:
\begin{enumerate}
    \item Detecting and removing the background of the image
    \item Removing dark grey periventricular areas along with other dark regions of the brain
    \item Segmenting bright areas of the brain image, which is mainly consisted of white matter lesions (if any)
\end{enumerate}
For each step, we perform adaptive thresholding based on the Otsu's threshold selection method \cite{otsu}. In each stage, the objective is to find the value of pixel intensity at which the between-class variance is maximized. This threshold value, $\theta$, divides the input image pixels into two groups $G_1$ and $G_2$:
\begin{equation}
    G_1 \leftarrow P~ \text{ if }~ 0 \leq I(P) \leq \theta ~~~ , ~~~ G_2 \leftarrow P \text{ if } ~ \theta < I(P) \leq 255,
\end{equation}
where $I(x)$ is the intensity value of the pixel $x$, and $P$ is one of the $m \times n$ pixels of the given gray scale image. 

\begin{figure}[t]
	\centering
	\includegraphics[width=1\textwidth]{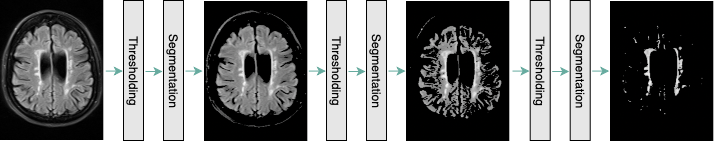}
	\caption{The proposed pre-processing method at a glance. Step 1: Removing background of the brain image; Step 2: Removing grey areas of the brain image, as they do not contain the white matter lesion; Step 3: Removing the remaining grey areas in the image and preserving the bright regions, which is mostly formed by White Matter Lesions.}
	\label{thresh}
\end{figure}

This process is conducted during each step, and the former output image is passed as the input image of the latter (Fig. \ref{thresh}). It is notable that since this method is performed on each image separately, it can adapt to images with different grayscale distributions. This enables us to segment white matter lesions in all patients regardless of the patient positioning and the device setting.

\subsection{Pre-training with Self-Supervision}
Pre-training the network with ImageNet \cite{deng2009imagenet,krizhevsky2012imagenet}, as a remedy for the small sample size problem, cannot be the appropriate approach due to marked differences in the structure of natural images and medical images \cite{tajbakhsh2016convolutional}. 
Self-supervised learning is an alternative strategy that extracts intrinsic information from the more abundant unlabeled data. It aims at supervised feature learning, where the supervision tasks are extracted from the data itself \cite{doersch2017multi,jing2020self}.

\begin{figure}[t]
     \centering
     \begin{subfigure}[t]{0.24\textwidth}
         \centering
         \includegraphics[width=\textwidth]{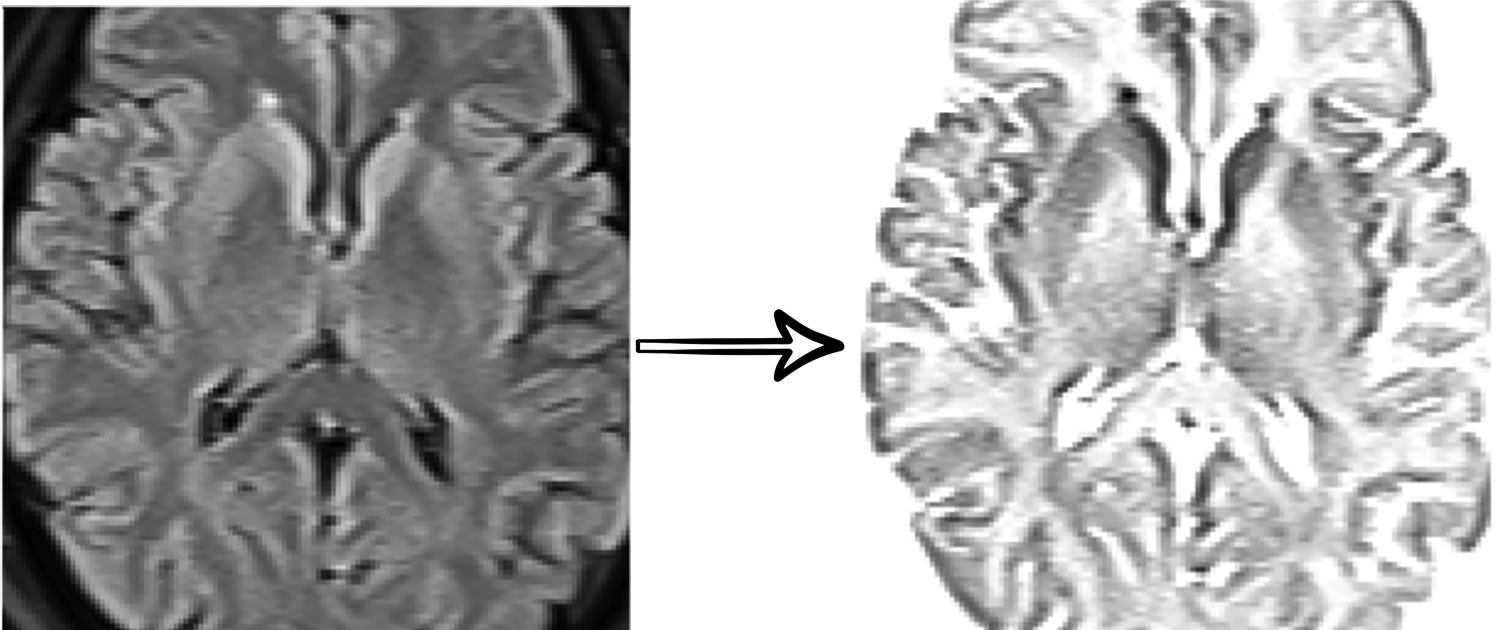}
         \caption{Non-linear transformation}
         \label{fig:transformation-non-linear}
     \end{subfigure}
     \hfill
     \begin{subfigure}[t]{0.24\textwidth}
         \centering
         \includegraphics[width=\textwidth]{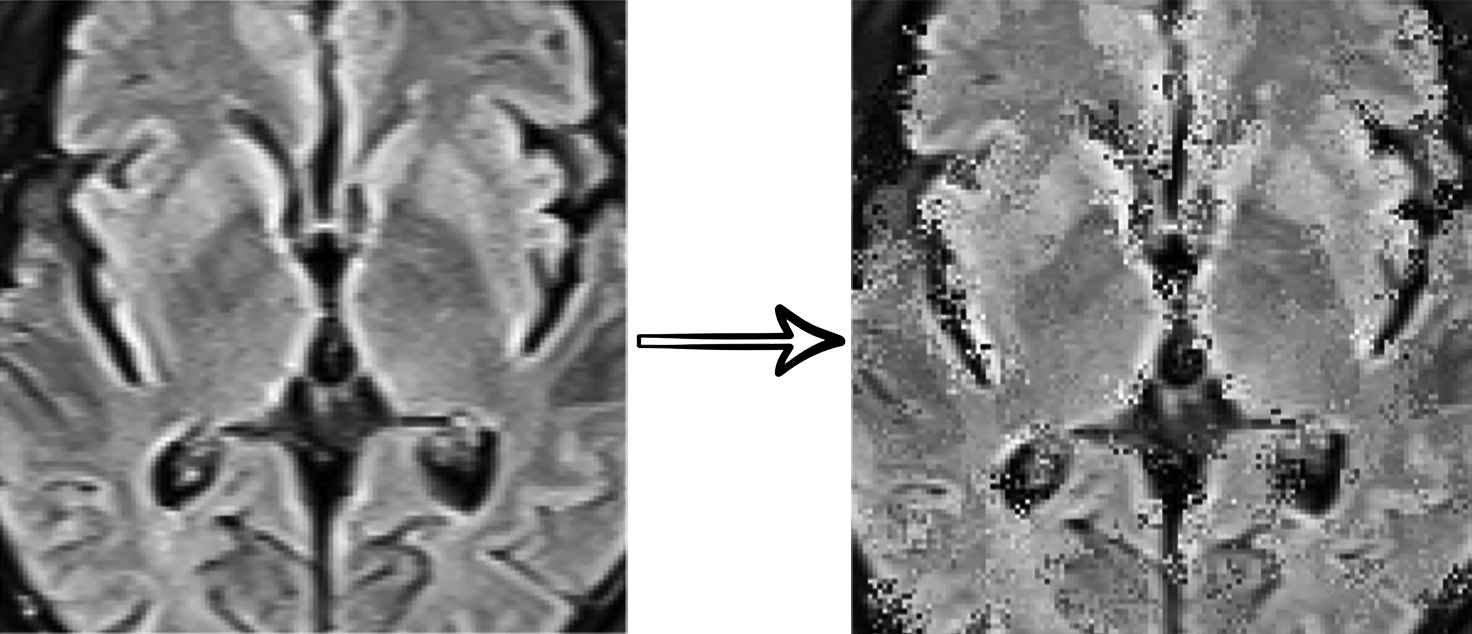}
         \caption{Local shuffling}
         \label{fig:transformation-local-shuffling}
     \end{subfigure}
     \hfill
     \begin{subfigure}[t]{0.24\textwidth}
         \centering
         \includegraphics[width=\textwidth]{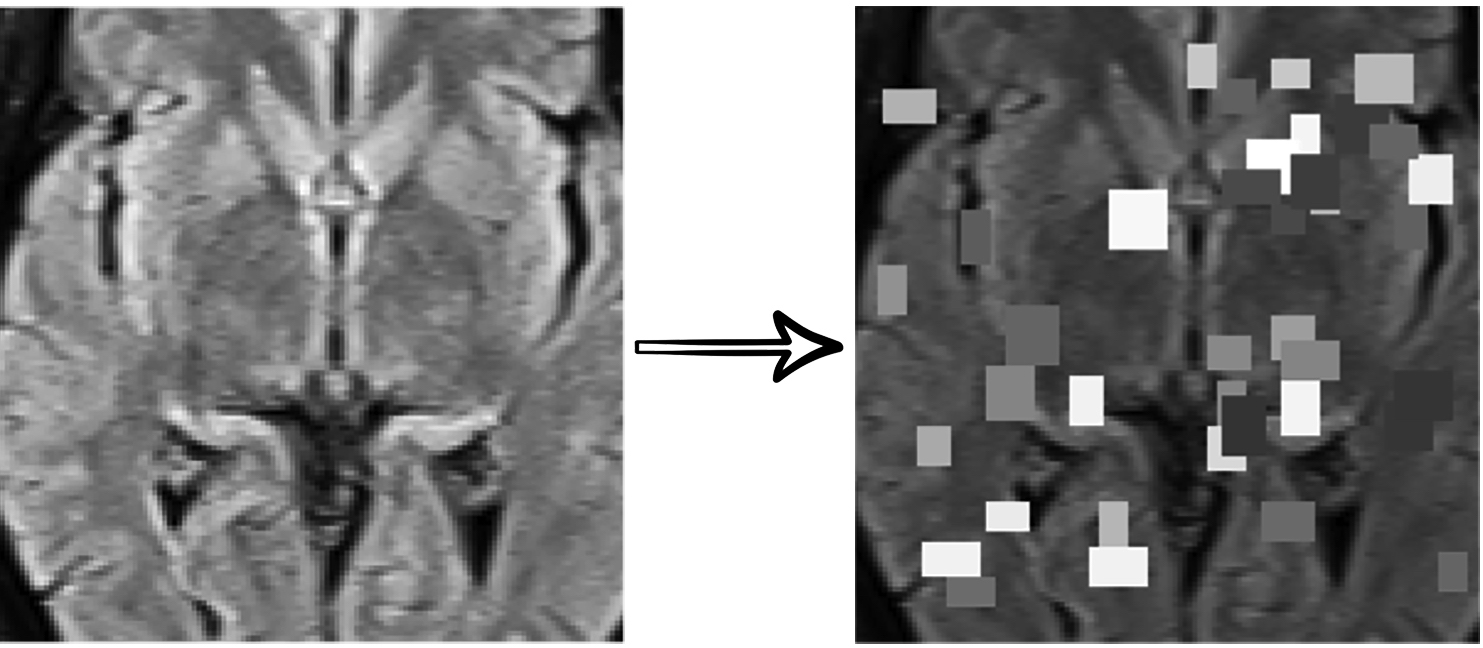}
         \caption{In-painting}
         \label{fig:transformation-in-painting}
     \end{subfigure}
     \hfill
     \begin{subfigure}[t]{0.24\textwidth}
         \centering
         \includegraphics[width=\textwidth]{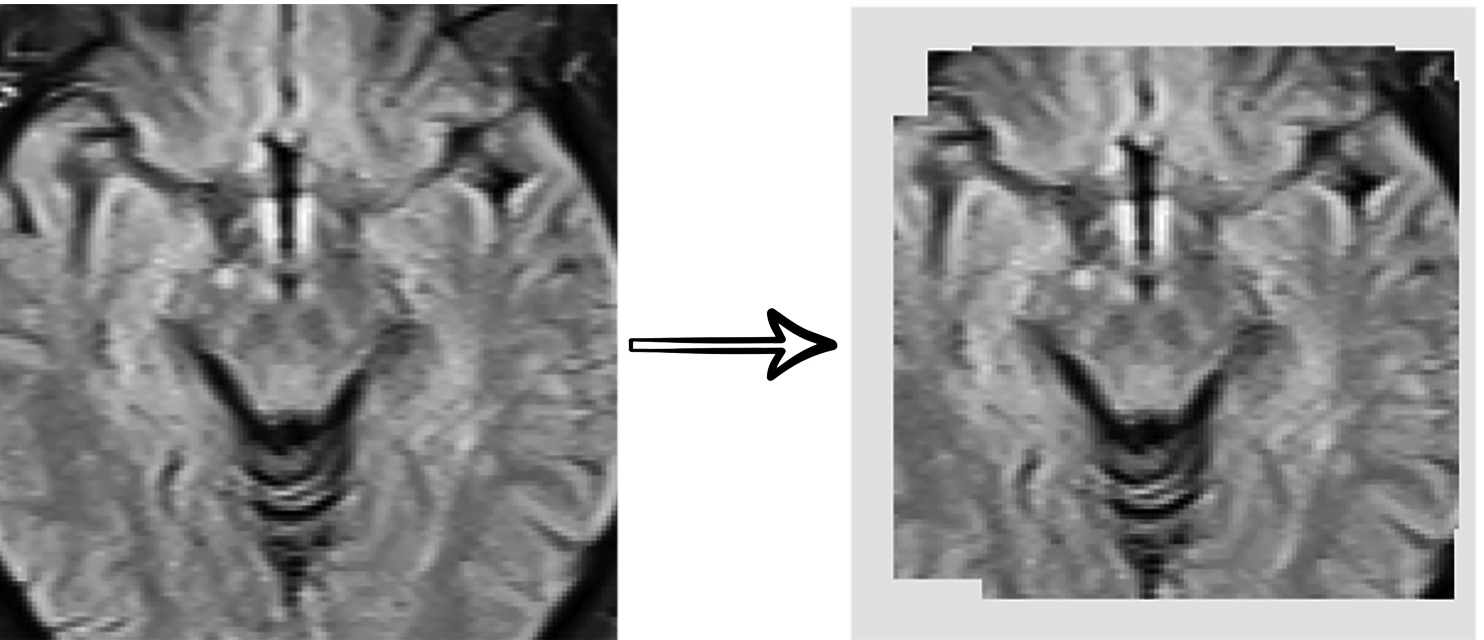}
         \caption{Out-painting}
         \label{fig:transformation-out-painting}
     \end{subfigure}
        \caption{Self-supervised transformations: \textbf{(a)} Non-linear Intensity Transformation enforces the model to learn the shape and intensity distribution as the appearance of organs.\textbf{(b)} Local Pixel Shuffling enforces the model to learn local boundaries as the texture of organs. \textbf{(c,d)} Out-painting and In-painting enforces the model to learn geometry and spatial layout of organs via extrapolating and local continuities of organs via interpolating as the context of images, respectively.}
        \label{fig:transformations}
\end{figure}

In this work, we use a self-supervised learning method to extract specific features of our data from multiple perspectives (appearance, texture, context, etc.) by combining three different supervision tasks that were proposed in ``Models Genesis" \cite{zhou2019models}. Our model consists of a U-Net \cite{ronneberger2015u} with a ResNet \cite{he2016deep} encoder. Each supervision task consists of a particular transformation and reconstructing the original image from the transformed one (see Fig. \ref{fig:transformations}). Finally, the encoder part of the network is fine-tuned for the target classification task.

\subsection{Attention Based Deep Multiple Instance Learning}
Unlike conventional supervised learning strategies in which a label exists for each training instance, in many medical image classification tasks, a single label is provided for a set of instances. 
Since we utilize thick-slice MRI data, we face the same weak supervision issue, so we use multiple instance learning (MIL).
 In a binary MIL problem, we assume there exists a label $Y_i \in \{ 0, 1\}$ for a bag of $K$ unordered instances $X_i = \{ x_{i}^{1}, \dots, x_{i}^{K} \}$,
each having its own label $y_{i}^{k}$ unknown to us.
MIL assigns a label to each bag based on the rule $Y_{i} = \max_{k}\{y_{i}^{k}\}$.

We propose to use an attention-based deep MIL network similar to \cite{ilse2018attention}.
As a relaxation to the original MIL problem, and to be able to apply gradient-based optimization methods, the bag label is modeled as a Bernoulli random variable, and its log-likelihood is optimized. Since the bag probability scoring function must be independent of the ordering of instances in each bag, it is symmetric and
can be decomposed in the following form \cite{ilse2018attention}:
\begin{equation}
    S(X) = g(\Sigma_{x \in X} f(x)),
\end{equation}
where $f$ and $g$ are suitable transformations and $\Sigma$ is a permutation-invariant operator, e.g., sum, weighted average, or max.
Therefore, our classification process consists of three steps: 
\begin{itemize}
    \item 
    Extract features from each instance using a function ($f$) which is modeled by a CNN with parameters $\theta$:
        \begin{equation}
            H = f_{\theta}(X)
        \end{equation}
    where $X \in \mathbb{R}^{K \times d \times d}$ is the data matrix and $H \in \mathbb{R}^{K \times u}$ is the extracted feature matrix.
    \item
    Make a permutation-invariant combination of the extracted features of all instances (a weighted average of features across each slice) using an operator which in our case is modeled by an attention mechanism network \cite{lin2017structured}:
        \begin{equation}
            A = \softmax(W_{s_{2}} \tanh(W_{s_{1}}H^{T})),
        \end{equation}
    where $W_{s_{1}} \in \mathbb{R}^{d_{a} \times u}$ and $W_{s_{2}} \in \mathbb{R}^ {r \times d_{a}}$ are arbitrary weight matrices and $d_a, r$ are hyperparameters. $A \in \mathbb{R}^{r \times K}$ is the annotations matrix.
    The final feature matrix thus becomes $M \in \mathbb{R}^{r \times u}$ computed by:
        \begin{equation}
            M = A H.
        \end{equation}
    \item
    Apply a final scoring function ($g$) on the combination to derive the bag label probabilities (the vector $Y$), modeled by a fully connected neural network. 
    \begin{equation}
        Y = g_{\phi}(M)
    \end{equation}
    
\end{itemize}


We extend the original work that is described in \cite{ilse2018attention}, and implement a multi-class MIL network. We replace the feature extraction CNN with the encoder of the U-Net (Fig. \ref{network-diagram}) to achieve better results. We consider the set of all original MRI slices of a subject together with their pre-processed form as a bag of instances. This enables us to interpret the attention mechanism neural network as a way of determining which slices are more influential on the output, and verifying our approach with the clinical findings.

\section{Evaluation}
\subsection{Dataset}
The private dataset that is used in this study is a collection of thick-slice MRI data. Considering the severity of damaged areas of the brain and their confluence, each patient's MRI is classified as follows: 0- absence, 1- punctuate foci, 2- smooth halo, and 3- large and irregular confluence areas of white matter changes in the brain. 
The data consists of 84 and 149 grade 1, 29 and 29 grade 2, 15 and 5 grade 3 dementia, and 172 and 117 healthy controls for PVWM and DWM biomarkers respectively, which makes a total of 300 subjects. 7 series of MRI has been acquired from each subject: 4 Axial series (T1-weighted, T2-weighted, PD, Flair), 1 Coronal series (T2-weighted), and 2 Sagittal series (T2-weighted, Flair). Each series contains 15-18 slices labeled at the patient level by a consensus of 3 experts. We exclusively utilized slices number 7-13 of the axial Flair series. Finally, it is worth noting that although the data that is employed in this study is remarkably imbalanced, it is in line with the distribution of the disease severity in the whole human society.

\subsection{Experimental Setup}
Since the dataset is severely imbalanced, we evaluate our model using the F1-Score. We randomly select 90\% of the data as the training set and the remaining data as the test set. We apply 10-fold cross-validation on the training data to perform a fair evaluation, and to select the best set of hyperparameters.
In order to tackle the issue of class imbalance, we use weighted sampling. We carefully augment the few data in hand by small random rotation and random horizontal flips. We ensure that there is no data leakage between the train and test sets by splitting the dataset at the subject level, i.e., slices from one subject do not appear in both the training and test datasets. Moreover, all data augmentation tasks are performed after splitting the dataset. 

A clear-cut comparison between dementia classification methods is intricate due to differences in datasets' populations, evaluation methodologies and the test distribution, labeling criteria, and the input type such as 2D slices, and full brain volume.\cite{mendoza2020single} Having that in mind, we decided to implement state-of-the-art methods, Gupta et al. \cite{gupta2013natural} and Islam et al. \cite{islam2018brain}, on popular public datasets, ADNI and OASIS, respectively. We filtered previous methods based on having 2D slices of the brain as their inputs and an independent subject test as their evaluation criteria. These filters are crucial in choosing state-of-the-art methods because these methodological differences have fundamental implications on the results' interpretation and comparison. We evaluated these two methods on our dataset to reach a fair comparison. We also provide the results for a VGG-16 \cite{DBLP:journals/corr/SimonyanZ14a} trained on the middle slice of each MRI data as a baseline CNN.

Ensemble of deep CNNs \cite{islam2018brain}, follows a particular connection pattern called dense connectivity. The main motivation behind dense connectivity, which was first proposed by \cite{huang2017densely}, is to reduce the overfitting in networks. This method consists of three deep convolutional neural networks, which are followed by a vote counter component. The sparse autoencoder \cite{gupta2013natural} contains a simple two-layer autoencoder that is pre-trained on natural images. The bases obtained during reconstructing natural images are used to extract medical image features and classify four levels of dementia based on these features.

\subsection{Results}
Table \ref{tab:component-analysis-table} summarizes the results of our method on the performance of the network over the visual biomarkers of the dementia dataset. Our approach, even without its pre-processing and self-supervision pre-training components, significantly outperforms the baseline VGG-16. Besides, adding each component alone improves the performance, with the influence of self-supervision pre-training being slightly higher. Promising results are obtained when both elements are in place. Table \ref{tab:component-analysis-table} also demonstrates the superiority of our method over \cite{gupta2013natural} and \cite{islam2018brain}. Investigation of the results of each biomarker separately suggests that the network generally performs better on the classification of PVWM changes due to the changes in DWM being less notable even in progressed stages of the disease. The severely imbalanced dataset leads to high standard deviations in the results, since in some cross-validation runs, there may be very few subjects from classes 2 and 3 in the training data.

\begin{table}[t]
\centering
\caption{Performance of the proposed method on the visual biomarkers of dementia dataset. 
Values correspond to the mean (and the standard deviation) of the macro averaged F1-score across 5 runs of 10-fold cross-validation}
\label{tab:component-analysis-table}
\begin{tabularx}{\textwidth}{@{}l@{\hskip 1.3cm}l@{\hskip 1cm}l@{}}
\toprule
                                       & \multicolumn{2}{l}{\textbf{macro averaged F1-score}}                 \\ \cmidrule(l){2-3} 
                                       & PVWM                   & DWM            \\ \midrule
\textbf{Baseline CNN}                           & $0.53 (0.034)$          & $0.49 4(0.047)$ \\
\textbf{Sparse Autoencoder \cite{gupta2013natural}}                           & $0.58 (0.133)$         & $0.541 (0.167)$ \\
\textbf{Ensemble system of deep CNNs \cite{islam2018brain}}                           & $0.61 (0.092)$          & $0.58(0.058)$ \\
\textbf{No Pre-processing, No Self-supervision} & $0.607 (0.123)$         & $0.628 (0.145)$ \\
\textbf{Pre-processing, No Self-supervision}    & $0.671 (0.111)$         & $0.637 (0.081)$ \\
\textbf{No Pre-processing, Self-supervision}    & $0.713 (0.110)$         & $0.689 (0.146)$ \\
\textbf{Pre-processing, Self-supervision}       & $0.76 (0.118)$          & $0.692 (0.074)$ \\ \bottomrule
\end{tabularx}
\end{table}

Table \ref{tab:class-accuracy} shows the quantitative results of the best settings of our method on the test data. The decrease in performance of the model on class $1$ of PVWM lesions is possibly due to insignificant changes to the brain in that class. Also, note that there are only 5 subjects with grade $3$ DWM changes, which has a major impact on the precision of our model.

\begin{table}[t]
\centering
\caption{Test results of the method on the dementia visual biomarkers dataset. Values correspond to the performance of the best configuration of the model with both pre-processing and self-supervision components.}
\label{tab:class-accuracy}
\begin{tabularx}{\textwidth}{@{}l@{\hskip 0.4cm}lll@{\hskip 0.2cm}lll@{\hskip 0.2cm}lll@{\hskip 0.2cm}ll@{}}
\toprule
\multirow{2}{*}{\textbf{\begin{tabular}[c]{@{}l@{}}Fazekas \\ Scale\end{tabular}}} & \multicolumn{2}{l}{\textbf{\begin{tabular}[c]{@{}l@{}}Relative \\ Frequency\end{tabular}}} &  & \multicolumn{2}{l}{\textbf{Precision}} &  & \multicolumn{2}{l}{\textbf{\begin{tabular}[c]{@{}l@{}}Sensitivity\\ (Recall)\end{tabular}}} &  & \multicolumn{2}{l}{\textbf{Specificity}} \\ \cmidrule(lr){2-3} \cmidrule(lr){5-6} \cmidrule(lr){8-9} \cmidrule(l){11-12} 
                                                                                   & \textbf{PVWM}                                & \textbf{DWM}                                &  & \textbf{PVWM}      & \textbf{DWM}      &  & \textbf{PVWM}                                 & \textbf{DWM}                                &  & \textbf{PVWM}       & \textbf{DWM}       \\ \midrule
\textbf{0}                                                                         & $57.33\%$                                   & $39\%$                                      &  & $0.94$             & $0.56$            &  & $0.81$                                        & $0.63$                                      &  & $0.89$              & $0.86$             \\
\textbf{1}                                                                         & $28\%$                                       & $49.67\%$                                   &  & $0.45$             & $0.75$            &  & $0.80$                                        & $0.75$                                      &  & $0.80$               & $0.83$             \\
\textbf{2}                                                                         & $9.67\%$                                     & $9.66\%$                                    &  & $1.00$             & $0.66$            &  & $0.67$                                        & $0.5$                                       &  & $1.00$              & $0.96$             \\
\textbf{3}                                                                         & $5\%$                                        & $1.67\%$                                    &  & $1.00$             & $0.33$            &  & $1.00$                                        & $1.00$                                      &  & $1.00$              & $0.93$             \\ \bottomrule
\end{tabularx}
\end{table}
    
\section{Conclusion}
We showed that white matter segmentation, self-supervised learning, and multiple instance learning provide an effective combination to deal with various challenges in dementia disease classification in the thick-slice MRI. These challenges include a small sample size, weak supervision, and inter-subject variations of the MRI. Our method outperforms state-of-the-art by large margins according to macro averaged F1-score. Due to the small sample size, we did not optimize various components deliberately to avoid overfitting. Therefore, such an optimization, which includes the choice of the loss function in MIL and region of interest encoding of the input, using a larger dataset is the subject of future work. 



\bibliographystyle{splncs04}


\bibliography{bibs/full,bibs/refs}
 
\newpage

\section{Appendix}
In this document, we provide additional materials to supplement our main submission. In the first section, the dataset which is provided and used for this paper is investigated. In the second section, we provide the mathematical details of Otsu Thresholding\cite{otsu}, which is used in pre-processing step. Third section illustrates four examples of our self-supervised training.

\subsection{Dataset}
MRI is an imaging technique that consists of a patch of sequential sheared 2D images of organs from 3 different orientation terms with various scan setups, leading to volumetric data. The basic orientation terms for an MRI of the body taken are Saggital, Coronal, and Axial. The most common MRI sequences are T1-weighted and T2-weighted scans. A third commonly used sequence is the Fluid Attenuated Inversion Recovery (Flair). The Flair sequence is similar to a T2-weighted image except that abnormalities are brighter and normal CSF fluid is attenuated and made dark. This sequence is very sensitive to pathology and makes the differentiation between CSF and an abnormality much easier. 
The MRI data are classified into two groups of thick-slice and thin-slice data. In thick-slice MRI, the distance between sections is higher, and consequently, the number of slices is lower. Capturing thick slice MRI is faster, making the patient less exposed to radiation. Experts mostly use thick-slice data to diagnose diseases. Considering the severity of damaged areas of the brain and their confluence, each patient's MRI is classified as follow: 0- absence, 1- punctuate foci, 2- smooth halo, and 3- large and irregular confluence areas of white matter changes in the brain (See Fig.\ref{fig:fazekas-scale})

\begin{figure}
     \centering
     \begin{subfigure}[t]{0.115\textwidth}
         \centering
         \includegraphics[width=\textwidth, height=0.09\textheight]{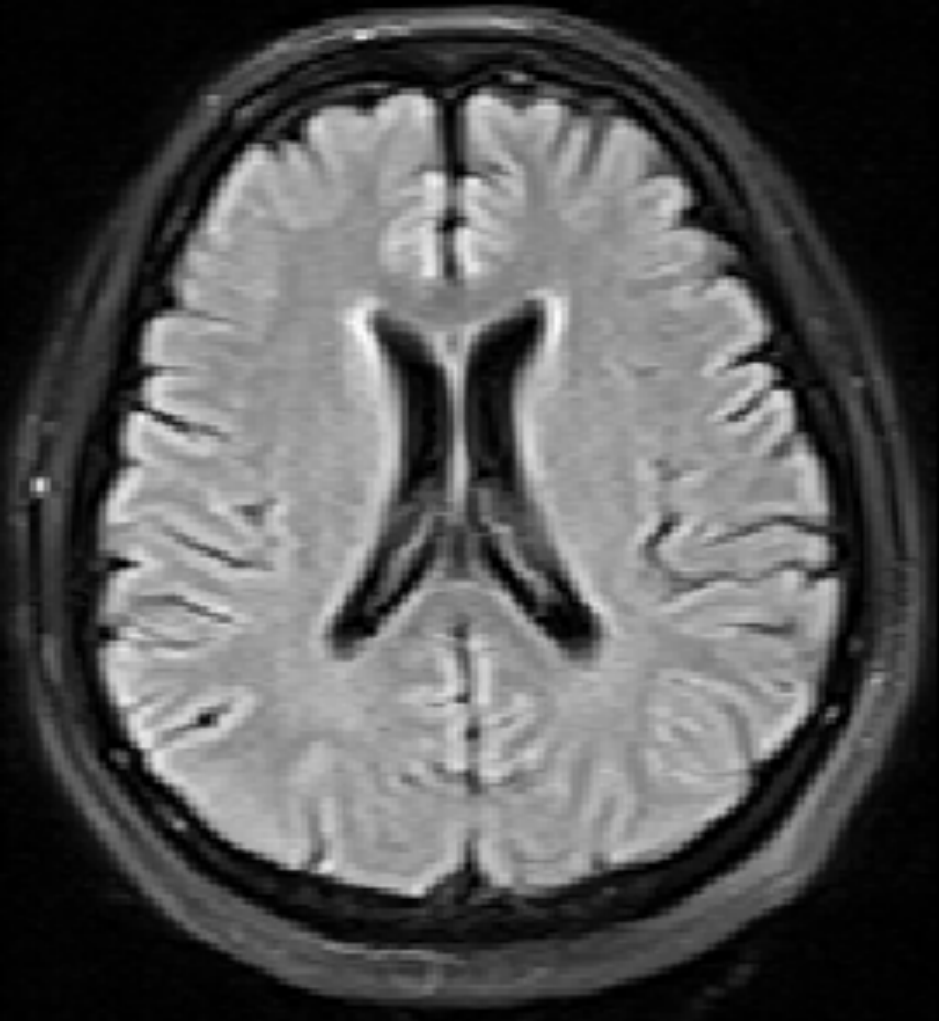}
         \caption{PVWM 0}
         \label{fig:fazekas-pvwm-0}
     \end{subfigure}
     \hfill
     \begin{subfigure}[t]{0.115\textwidth}
         \centering
         \includegraphics[width=\textwidth, height=0.09\textheight]{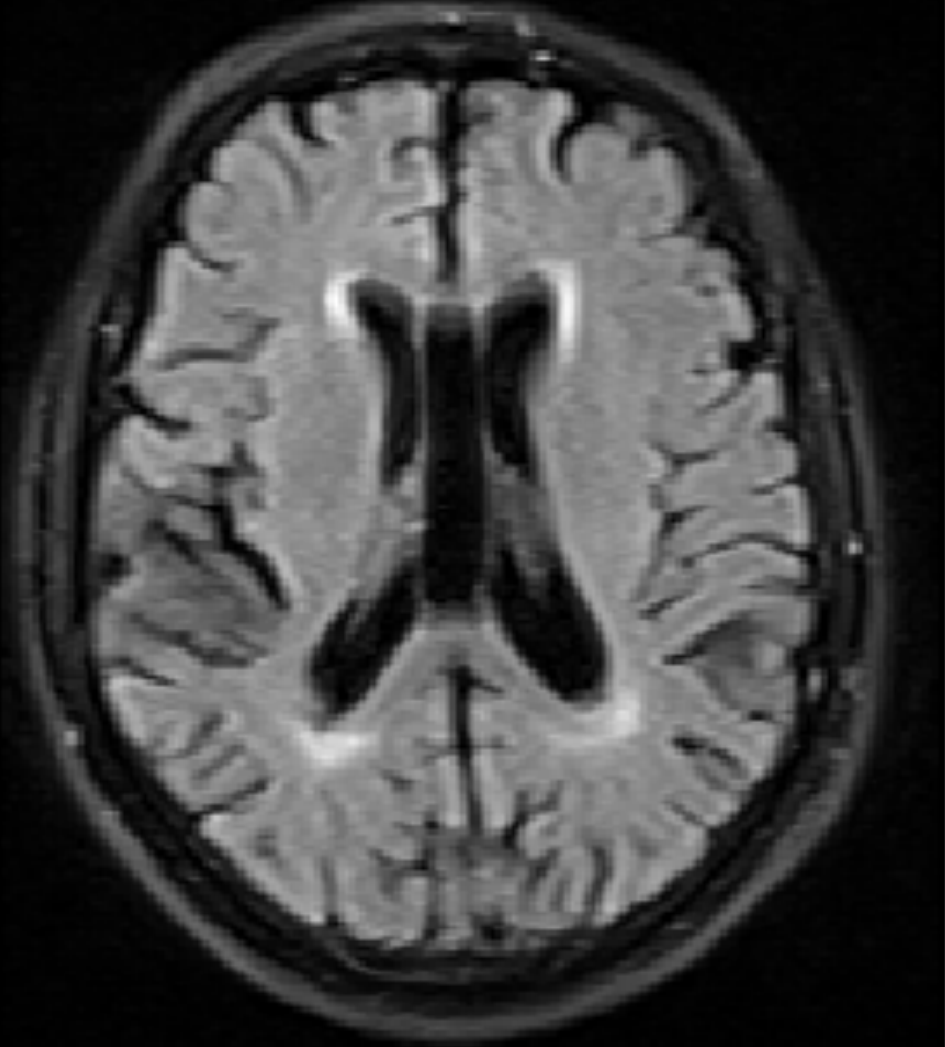}
         \caption{PVWM 1}
         \label{fig:fazekas-pvwm-1}
     \end{subfigure}
     \hfill
     \begin{subfigure}[t]{0.115\textwidth}
         \centering
         \includegraphics[width=\textwidth, height=0.09\textheight]{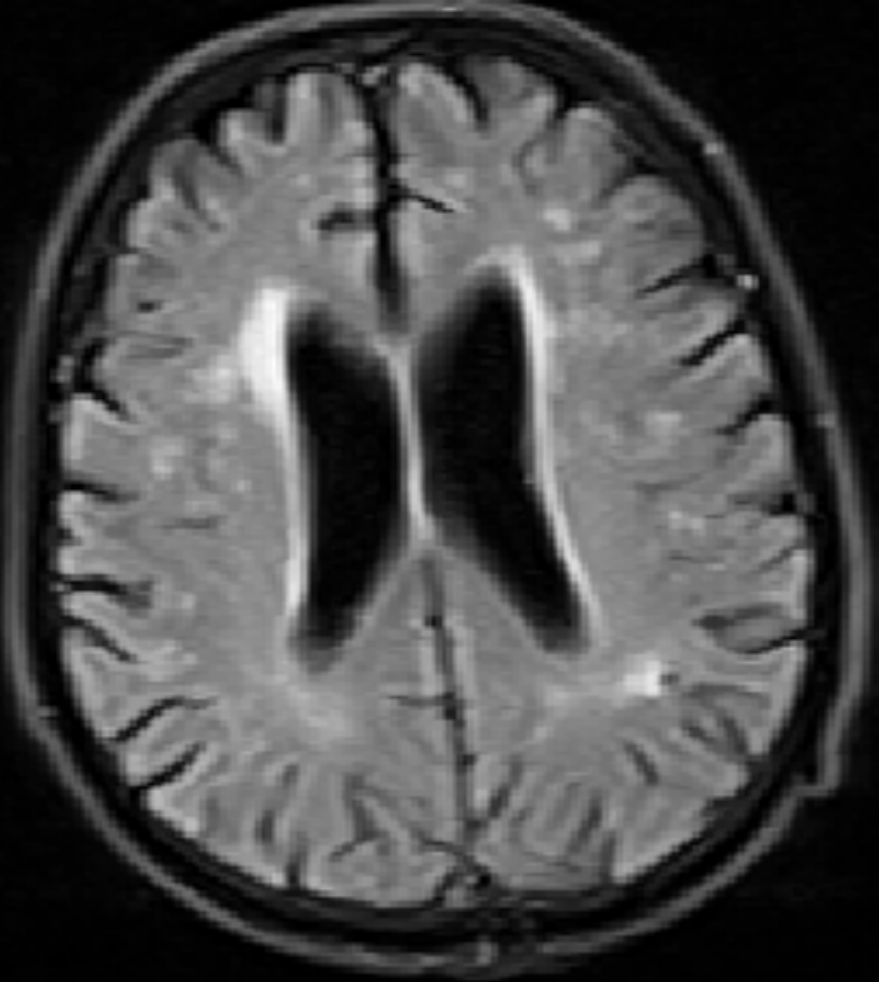}
         \caption{PVWM 2}
         \label{fig:fazekas-pvwm-2}
     \end{subfigure}
     \hfill
     \begin{subfigure}[t]{0.115\textwidth}
         \centering
         \includegraphics[width=\textwidth, height=0.09\textheight]{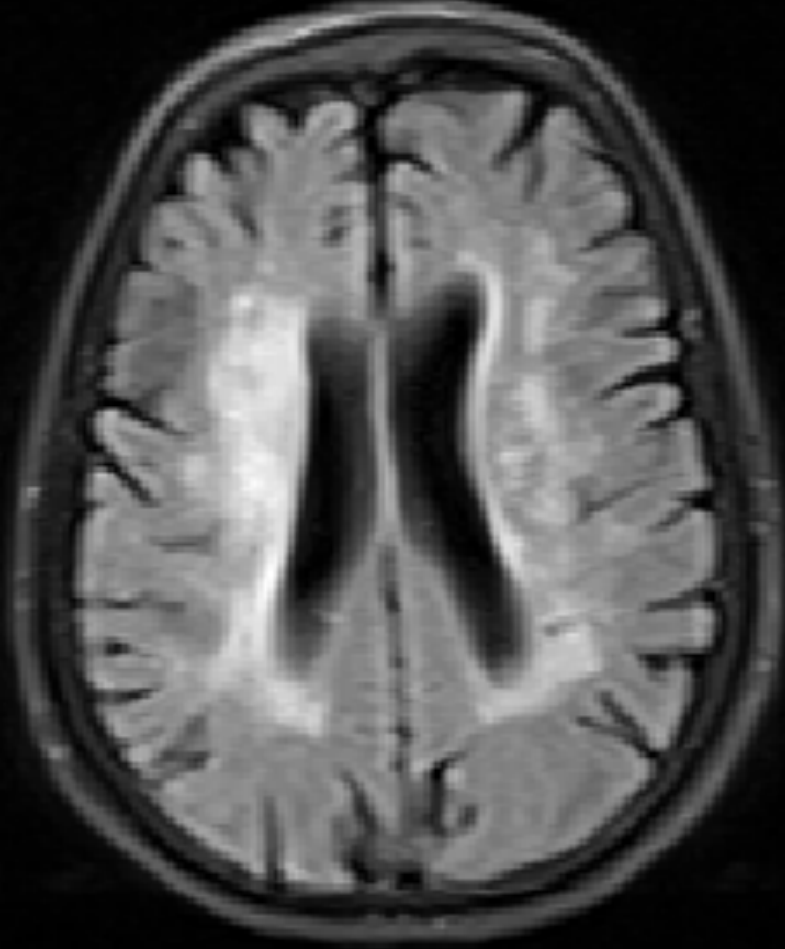}
         \caption{PVWM 3}
         \label{fig:fazekas-pvwm-3}
     \end{subfigure}
     \begin{subfigure}[t]{0.115\textwidth}
         \centering
         \includegraphics[width=\textwidth, height=0.09\textheight]{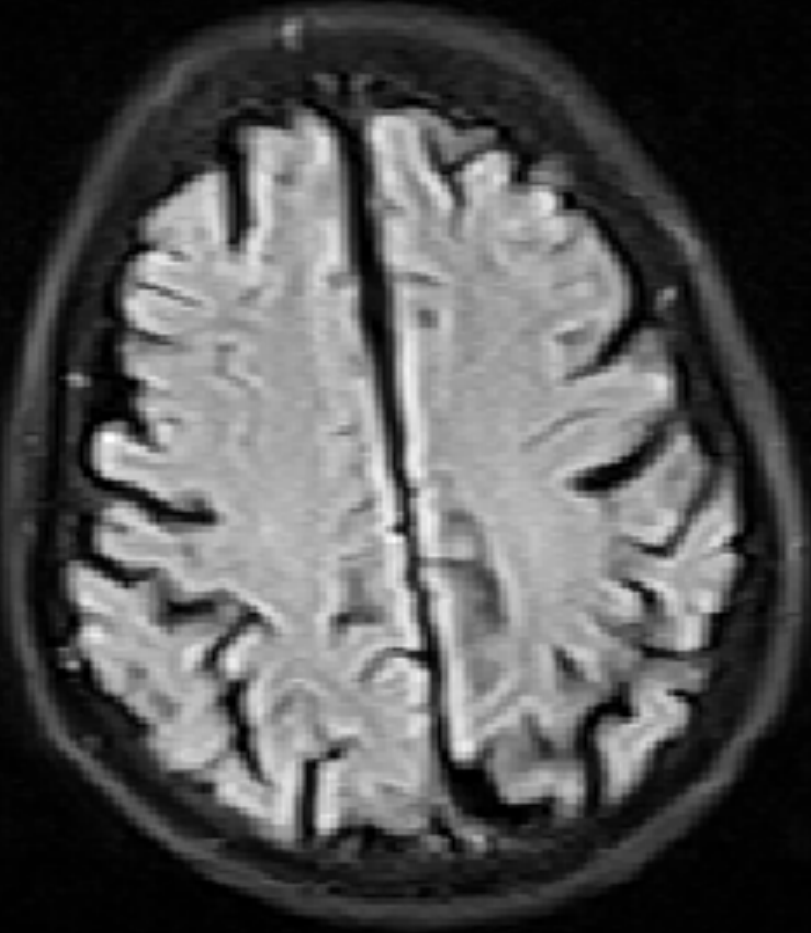}
         \caption{DWM 0}
         \label{fig:fazekas-dwm-0}
     \end{subfigure}
     \hfill
     \begin{subfigure}[t]{0.115\textwidth}
         \centering
         \includegraphics[width=\textwidth, height=0.09\textheight]{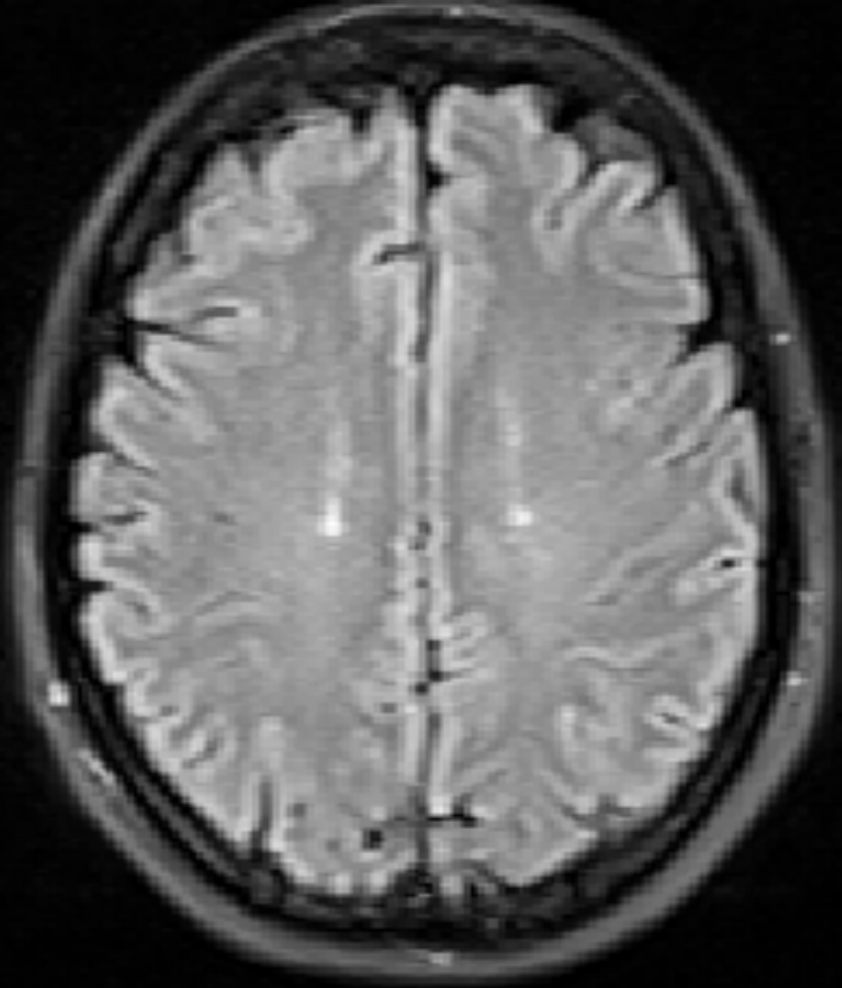}
         \caption{DWM 1}
         \label{fig:fazekas-DWM-1}
     \end{subfigure}
     \hfill
     \begin{subfigure}[t]{0.115\textwidth}
         \centering
         \includegraphics[width=\textwidth, height=0.09\textheight]{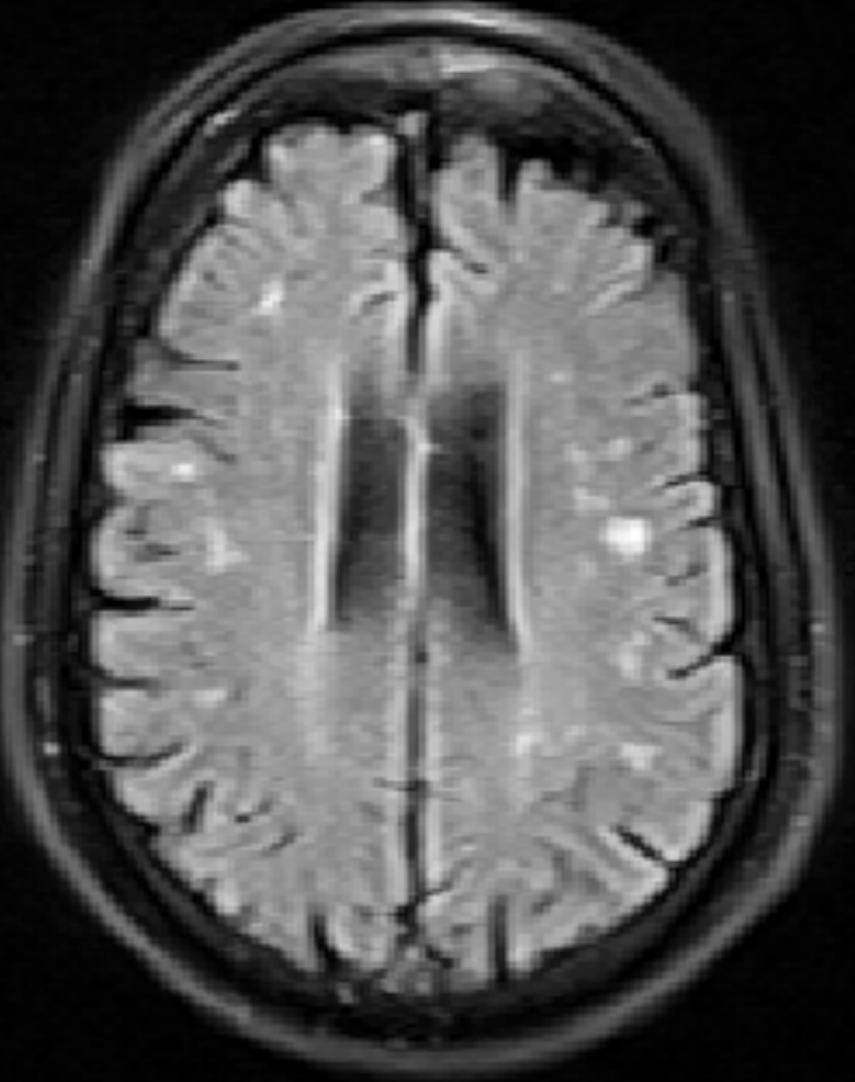}
         \caption{DWM 2}
         \label{fig:fazekas-dwm-2}
     \end{subfigure}
     \hfill
     \begin{subfigure}[t]{0.115\textwidth}
         \centering
         \includegraphics[width=\textwidth, height=0.09\textheight]{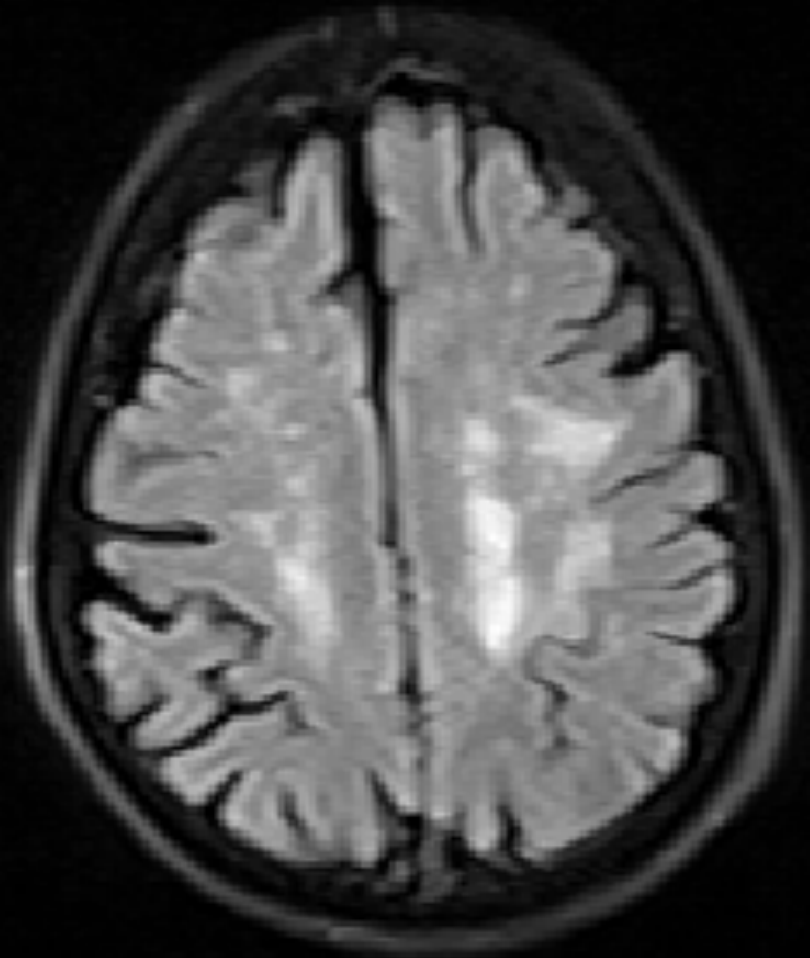}
         \caption{DWM 3}
         \label{fig:fazekas-dwm-3}
     \end{subfigure}
        \caption{Fazekas scale for the DWM and PVWM lesions. Although the grade of the biomarkers are correlated, they are not necessarily equal for the same subject. Also note that the PVWM changes are more remarkable than DWM changes.}
        \label{fig:fazekas-scale}
\end{figure}

\subsection{Otsu Thresholding}
For optimizing the threshold value, as proposed an exhaustive search is used for computing the threshold value such that it minimizes the intra-class variance, which is a weighted sum of variances of both classes:
\begin{equation} \label{eqotsu}
\sigma^2_\omega(t) = \omega_0(t)\sigma^2_0(t) + \omega_1(t)\sigma^2_1(t).
\end{equation}
The terms $\sigma^2_0$ and $\sigma^2_1$ are the two classes' variances, and $\omega_c$ is the probability of the class occurrence that can be computed by
\begin{equation}
    \omega_c(t) = \sum_{i \in c} \frac{n_i}{N}.
\end{equation}
Here, $i$ is the pixel intensity that is in class $c$, $n_i$ is the number of pixels in the given image that has the intensity of $i$, and $N$ is the total number of pixels. On the other hand, the class variance is computed by
\begin{equation}
    \sigma_c^2(t)=\sum_{i \in c} \frac{(i-\mu_c)^2p_i}{\omega_c}
\end{equation}
where $\mu_c$ is the class mean, and $p_i$ is the probability of pixel intensity occurrence, $n_i/N$. By minimizing Eq.\ref{eqotsu}, the threshold value is computed, and pixels are classified into two groups, where the $G_1$ contains dark pixels and the $G_2$ contains bright ones. Finally, each pixel's intensity value in $G_1$ is replaced with 0, which is the darkest color representation in grayscale images.\\

\subsection{Self-Supervised Training Results Visualization}

\begin{figure}
     \centering
     \includegraphics[width=0.9\textwidth]{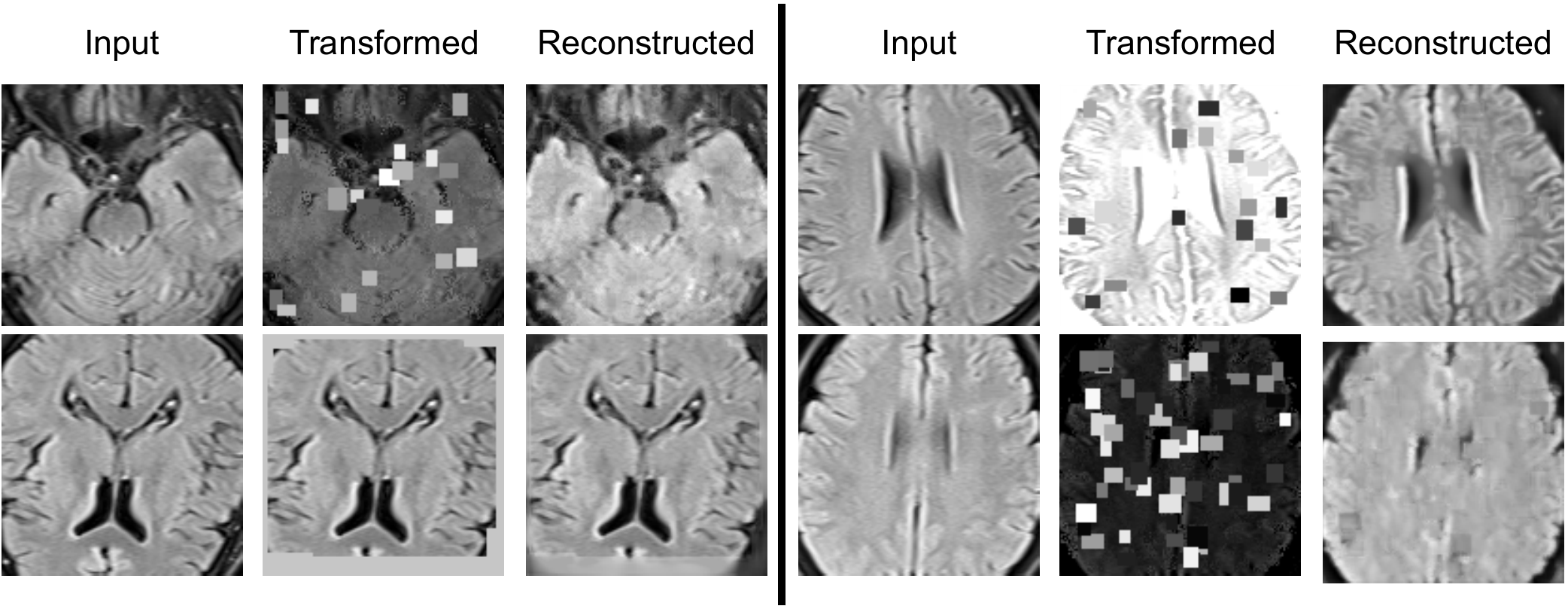}
        \caption{ Reconstruction results of the combination of various transformations over images through the self-supervised network.}
        \label{fig:fazekas-scale}
\end{figure}


\end{document}